\documentclass[aps,prd,preprint,superscriptaddress,showpacs]{revtex4}

\usepackage{graphicx}
\usepackage{ulem}
\usepackage{color}
\definecolor{My_red}        {cmyk}{0.00,1.00,1.00,0.20}

\def\ie{{\it i.e.}}

\def\nnb{\nonumber}

\def\bwt{\begin{widetext}}
\def\ewt{\end{widetext}}
\def\be{\begin{equation}}
\def\ee{\end{equation}}
\def\bea{\begin{eqnarray}}
\def\eea{\end{eqnarray}}
\def\bean{\begin{eqnarray*}}
\def\eean{\end{eqnarray*}}
\def\bary{\begin{array}}
\def\eary{\end{array}}
\def\bit{\begin{itemize}}
\def\eit{\end{itemize}}
\def\GeV{\rm GeV}

\def\su5u1{SU(5) \times U(1)}
\def\fsu5u1{SU(5) \times U(1)'}
\def\so10{SO(10)}
\def\sq20{SO(10) \times SO(10)}

\begin{document}

\title{Testable Flipped $SU(5)\times U(1)_X$ Models}

\author{Jing Jiang}
\affiliation{Institute of Theoretical Science, University of Oregon, 
Eugene, OR 97403, USA}

\author{Tianjun Li}

\affiliation{George P. and Cynthia W. Mitchell Institute for
Fundamental Physics, Texas A$\&$M University, College Station, TX
77843, USA }

\affiliation{Institute of Theoretical Physics, Chinese Academy of Sciences,
 Beijing 100080, P. R. China}

\affiliation{Department of Physics and Astronomy, Rutgers University, 
Piscataway, NJ 08854, USA}

\author{Dimitri V. Nanopoulos}

\affiliation{George P. and Cynthia W. Mitchell Institute for
Fundamental Physics, Texas A$\&$M University, College Station, TX
77843, USA }

\affiliation{Astroparticle Physics Group, Houston Advanced
Research Center (HARC), Mitchell Campus, Woodlands, TX 77381, USA}

\affiliation{Academy of Athens, Division of Natural Sciences, 28
Panepistimiou Avenue, Athens 10679, Greece }

\date{\today}

\begin{abstract}

The little hierarchy between the GUT scale and the string scale 
may give us some hints that can be tested at the LHC. To achieve 
 string-scale gauge coupling unification, we introduce 
additional vector-like particles. We require that these vector-like
particles be standard, form complete GUT multiplets, and
have masses around the TeV scale or close to the string scale.
Interestingly, only the flipped $SU(5)\times U(1)_X$ models 
can work elegantly. 
We consider all possible sets of vector-like particles
with masses around the TeV scale. And we introduce vector-like
particles with masses close to the string scale which can mimic
the string-scale threshold corrections. We emphasize that
all of these vector-like particles can be obtained in the 
interesting flipped $SU(5)\times U(1)_X$ string models 
from the four-dimensional free fermionic string construction. 
Assuming the low-energy supersymmetry, high-scale supersymmetry,
and  split supersymmetry, we show that the string-scale
gauge coupling unification can indeed be achieved in the
flipped $SU(5)\times U(1)_X$ models. These models can
be tested at the LHC by observing simple sets of
vector-like particles at the TeV scale. Moreover,
we  discuss a simple flipped $SU(5)\times U(1)_X$ model 
with string-scale gauge coupling unification and
high-scale supersymmetry by introducing only
one pair of the vector-like particles at the
TeV scale, and we predict the corresponding Higgs boson masses.
Also, we briefly comment on the string-scale gauge coupling
unification in the model with low-energy supersymmetry 
by introducing only one pair of the vector-like
particles at the intermediate scale. And we briefly comment
on the mixings among the SM fermions and the corresponding extra
vector-like particles.

\end{abstract}


\preprint{ACT-07-06, MIFP-06-25, OITS-786, RUNHETC-06-24, hep-ph/0610054}

\maketitle


\section{Introduction}

Supersymmetry provides an elegant solution to the gauge
hierarchy problem, and Grand Unified Theories (GUTs) gives us a
simple understanding of the quantum numbers of the Standard Model
(SM) fermions. In particular, the success of gauge coupling
unification in the Minimal Supersymmetric Standard Model (MSSM)
strongly supports the possibility of the supersymmetric 
GUTs~\cite{Langacker:1991an}.  
Moreover, the electroweak gauge symmetry can be broken 
by the radiative corrections due to the large top quark
Yukawa coupling~\cite{Ellis:1982wr}, and  the tiny neutrino masses can be 
generated naturally via the see-saw mechanism.
 Therefore, supersymmetric GUTs may describe all the known fundamental
interactions in nature except gravity.

The interesting question is whether we can test
 supersymmetric GUTs at future colliders and
experiments. The major prediction of supersymmetric 
GUTs is the dimension-5 proton decay from the 
colored Higgsino exchange~\cite{Enqvist:1984vs}.
This kind of proton decay is
suppressed due to the Yukawa couplings. However,  we can
introduce the non-renormalizable operators to mimic
such proton decay, {\it i.~e.}, generic
dimensional-5 proton decay opertors with
Planck scale suppression and without Yukawa coupling
suppression~\cite{Ellis:1983qm}. So, 
even if we observe such proton decay at future
experiments, we can not confirm the possibility of 
supersymmetric GUTs.

If string theory is correct, 
it seems to us that  one new clue is the little hierarchy
between the GUT scale $M_{\rm GUT}$ and the string scale $M_{\rm string}$.
It is well-know that the gauge coupling unification scale in
the MSSM, which is called the GUT scale in
the literature, is around $2\times 10^{16}$ GeV~\cite{Langacker:1991an}.
The gauge coupling unification in the MSSM
 is based on two implicit assumptions: (1)
the $U(1)_Y$ normalization is canonical; (2)  there are no intermediate-scale 
threshold corrections. On the other hand, 
 the string scale $M_{\rm string}$ in the weakly coupled heterotic 
string theory is~\cite{Dienes:1996du}
\begin{eqnarray}
M_{\rm string} = g_{\rm string} \times 5.27 \times 10^{17} ~{\rm GeV}~,~\,
\end{eqnarray}
where $g_{\rm string}$ is the string coupling constant. Note that since
$g_{\rm string} \sim {\cal O} (1)$, we have
\begin{eqnarray}
M_{\rm string} \approx 5 \times 10^{17} ~{\rm GeV}~.~\,
\end{eqnarray}
Thus, there exists a factor of approximately 20 to 25 between 
 the MSSM unification scale and the string scale (In the
strong coupled heterotic string theory or M-theory 
on $S^1/Z_2$~\cite{Horava:1995qa},
the eleven-dimensional Planck scale can be the MSSM unification 
scale due to the large eleventh dimension~\cite{Witten:1996mz}.
But  in this paper we concentrate on the weakly coupled heterotic 
string theory.). The discrepancy between the scales $M_{\rm GUT}$
and $M_{\rm string}$ implies that the weakly coupled heterotic 
string theory naively predicts wrong values for the
electroweak mixing angle ($\sin^2\theta_W$) and strong coupling
($\alpha_3$) at the weak scale. Because the weakly coupled heterotic 
string theory is one of the leading candidates for
a unified theory of the fundamental particles and interactions
in the nature, how to achieve the string-scale gauge coupling unification
is an important question in string 
phenomenology~\cite{Kaplunovsky:1987rp, Dixon:1990pc, Antoniadis:1991cf, Ibanez:1993bd,
Dienes:1995sv, Mayr:1995rx, Nilles:1995kb, Martin:1995wb, Bachas:1995yt, 
Bastero-Gil:1999dx, Giedt:2002kb, Emmanuel-Costa:2005nh, BJLL-SU}.

In general, we can always achieve the string-scale gauge coupling
unification by introducing additional vector-like particles with arbitrary
masses or arbitrary SM quantum numbers. Therefore, in order to have  interesting
and natural models, we make the following three requirements for the
additional vector-like particles: \\

(1) All the vector-like particles are standard.
We define the standard particles as the particles that can decay into
the MSSM particles via Yukawa couplings. So, the extra
vector-like particles are not stable, and then there are no
strong cosmological constraints on them. \\

(2) All the vector-like particles  must form the 
complete GUT multiplets. So, we do not need to split the
multiplets, which is a generic problem in the GUTs. \\

(3)  All the vector-like particles must have masses
around the TeV scale or close to the string scale, and
the string-scale threshold corrections can not be very large
which is  unnatural. So, the TeV-scale vector-like particles can be 
produced and tested at the Large Hadron Collider (LHC),
and the superheavy vector-like particles can be considered as
the string-scale threshold corrections. Also, there
are no intermediate-scale threshold corrections.\\

From requirement (2), we can not achieve the string-scale gauge
coupling unification in all the GUTs with simple GUT groups, 
for example, $SU(5)$, $SO(10)$ and $E_6$. 
Also, in the MSSM or standard-like supersymmetric Standard Models
that can be constructed from the weakly coupled heterotic 
string theory directly, to  achieve the $SU(2)_L$ and
 $SU(3)_C$ gauge coupling unification at the
string scale, we either need intermediate-scale ($10^{13}~{\rm GeV}$)
  threshold corrections or very large string-scale threshold
corrections~\cite{BJLL-SU}, which violates the requirement (3).
In addition,
  to achieve string-scale gauge coupling unification, we should
  introduce at the TeV scale sets of vector-like particles in GUT
  multiplets, whose contribution to the one-loop beta functions of the
  $U(1)_Y$, $SU(2)_L$ and $SU(3)_C$ gauge symmetry,  $\Delta b_1$,
  $\Delta b_2$ and $\Delta b_3$ respectively, satisfy $\Delta b_1 <
  \Delta b_2 = \Delta b_3$.
Since there is no such set of vector-like particles 
forming complete GUT representations
in the Pati-Salam models with symmetric $SU(2)_L$ and $SU(2)_R$,
  we can not achieve the string-scale gauge coupling
unification with the above requirements. Furthermore, in 
the Pati-Salam models with asymmetric $SU(2)_L$ and $SU(2)_R$,
similar to the MSSM or
 standard-like supersymmetric Standard Models
we can not achieve the string-scale gauge coupling unification
for $SU(2)_L\times SU(3)_C$ or $SU(2)_L\times SU(4)$
unless there exist intermediate-scale or very large
 string-scale threshold corrections. Thus, in this paper 
we will not consider the MSSM, standard-like supersymmetric 
Standard Models, and Pati-Salam models with
 asymmetric $SU(2)_L$ and $SU(2)_R$, although these
models can be constructed in the weakly coupled
heterotic string theory~\cite{Faraggi:1989ka, Antoniadis:1990hb,
Faraggi:1991jr, Chaudhuri:1995ve, Cleaver:1997jb, Leontaris:1995sf}.

Interestingly enough, in the flipped $SU(5)\times U(1)_X$ 
models~\cite{smbarr, dimitri, AEHN-0, AEHN, LNY, Huang:2003fv, Kim:2006hv},
we do have such kind of vector-like particles, for instance, 
the $XF$ and $\overline{XF}$ with respectively the quantum numbers
${\mathbf{(10, 1)}}$ and ${\mathbf{({\overline{10}}, -1)}}$
under the $SU(5)\times U(1)_X$ gauge symmetry.
Especially, flipped $SU(5)\times U(1)_X$ 
models can be constructed naturally in the weakly coupled 
heterotic string theory 
at Kac-Moody level one~\cite{AEHN, LNY, Kim:2006hv}.
Therefore, with the requirements above, we can only
achieve the string-scale gauge coupling unification
in the flipped $SU(5)\times U(1)_X$ models elegantly.
In fact, introducing such vector-like particles with
masses in the intermediate scale, the 
string-scale gauge coupling unification has been realized
previously~\cite{Lopez:1993qn, Lopez:1995cs}.

We also require that all the gauge couplings have
no Landau pole problem below the string scale.
To systematically study the string-scale gauge coupling
unification, we consider all the possible sets of vector-like particles
with masses around the TeV scale. And we introduce vector-like
particles with masses close to the string scale which can mimic
the string-scale threshold corrections. We emphasize that
all of these vector-like particles can be obtained in the 
interesting flipped $SU(5)\times U(1)_X$ string models from the
four-dimensional free fermionic formulation of
the weakly coupled heterotic string theory~\cite{LNY}.
Moreover, for the supersymmetry breaking scenarios,
we will consider the low energy supersymmetry, 
high-scale supersymmetry~\cite{Barger:2004sf, Barger:2005gn}, and 
 split supersymmetry~\cite{NASD, Giudice:2004tc}.
We will show that the string-scale gauge coupling unification
can indeed be realized. For the high-scale supersymmetry 
and split supersymmetry, we also calculate the corresponding
Higgs boson masses.

Furthermore, we briefly discuss a simple flipped 
$SU(5)\times U(1)_X$ model 
with string-scale gauge coupling unification and
high-scale supersymmetry breaking by introducing only
one pair of the vector-like particles at the
TeV scale, and we predict the Higgs boson masses.
Also, we briefly comment on a simple model with low-energy
supersymmetry and one pair of intermediate-scale vector-like
particles. And we briefly comment
on the mixings among the SM fermions and the corresponding extra
vector-like particles.

This paper is organized as follows: in Section II, we briefly
review the flipped $SU(5)\times U(1)_X$ models and the calculations of
Higgs boson mass in the high-scale supersymmetry and 
split supersymmetry. We study the string-scale gauge coupling
unification in the flipped $SU(5)\times U(1)_X$ models
in Section III. We  discuss simple  
flipped $SU(5)\times U(1)_X$ models with string-scale 
gauge coupling unification in Section IV. 
We comment on the mixings among the SM fermions and the corresponding extra
vector-like particles in Section V. Our discussion and
conclusions are in Section VI. The renormalization
group equations (RGEs) and beta functions 
for the non-supersymmetric and supersymmetric Standard
Models with additional vector-like particles
are given in Appendix A, and for the flipped $SU(5)\times U(1)_X$
models with additional vector-like particles are given in Appendix B.

\section{Brief Review}

\subsection{Flipped $SU(5)$ Model}

In this subsection, we would like to briefly review the flipped
$SU(5)$ model~\cite{smbarr, dimitri, AEHN-0}. 
The gauge group for flipped $SU(5)$ model is
$SU(5)\times U(1)_{X}$, which can be embedded into $SO(10)$ model.
We define the generator $U(1)_{Y'}$ in $SU(5)$ as 
\bea 
T_{\rm U(1)_{Y'}}={\rm diag} \left(-{1\over 3}, -{1\over 3}, -{1\over 3},
 {1\over 2},  {1\over 2} \right).
\label{u1yp}
\eea
The hypercharge is given by
\bea
Q_{Y} = {1\over 5} \left( Q_{X}-Q_{Y'} \right).
\label{ycharge}
\eea

There are three families of the SM fermions 
whose quantum numbers under $SU(5)\times U(1)_{X}$ are
\bea
F_i={\mathbf{(10, 1)}},~ {\bar f}_i={\mathbf{(\bar 5, -3)}},~
{\bar l}_i={\mathbf{(1, 5)}},
\label{smfermions}
\eea
where $i=1, 2, 3$.
As an example, the particle assignments for the first family are
\bea
F_1=(Q_1, D^c_1, N^c_1),~{\overline f}_1=(U^c_1, L_1),~{\overline l}_1=E^c_1~,~
\label{smparticles}
\eea
where $Q$ and $L$ are respectively the superfields of the left-handed
quark and lepton doublets, $U^c$, $D^c$, $E^c$ and $N^c$ are the
$CP$ conjugated superfields for the right-handed up-type quark,
down-type quark, lepton and neutrino, respectively,
To generate the heavy right-handed neutrino masses, we introduce
three SM singlets $\phi_i$.

To break the GUT and electroweak gauge symmetries, we introduce two pairs
of Higgs representations
\bea
H={\mathbf{(10, 1)}},~{\overline{H}}={\mathbf{({\overline{10}}, -1)}},
~h={\mathbf{(5, -2)}},~{\overline h}={\mathbf{({\bar {5}}, 2)}}.
\label{Higgse1}
\eea
We label the states in the $H$ multiplet by the same symbols as in
the $F$ multiplet, and for ${\overline H}$ we just add ``bar'' above the fields.
Explicitly, the Higgs particles are
\bea
H=(Q_H, D_H^c, N_H^c)~,~
{\overline{H}}= ({\overline{Q}}_{\overline{H}}, {\overline{D}}^c_{\overline{H}}, 
{\overline {N}}^c_{\overline H})~,~\,
\label{Higgse2}
\eea
\bea
h=(D_h, D_h, D_h, H_d)~,~
{\overline h}=({\overline {D}}_{\overline h}, {\overline {D}}_{\overline h},
{\overline {D}}_{\overline h}, H_u)~,~\,
\label{Higgse3}
\eea
where $H_d$ and $H_u$ are one pair of Higgs doublets in the MSSM.
We also add one singlet $S$.

To break the $SU(5)\times U(1)_{X}$ gauge symmetry down to the SM
gauge symmetry, we introduce the following Higgs superpotential at the GUT scale
\bea
{\it W}_{\rm GUT}=\lambda_1 H H h + \lambda_2 {\overline H} {\overline H} {\overline
h} + S ({\overline H} H-M_{\rm H}^2). 
\label{spgut} 
\eea 
There is only
one F-flat and D-flat direction, which can always be rotated along
the $N^c_H$ and ${\overline {N}}^c_{\overline H}$ directions. So, we obtain that
$<N^c_H>=<{\overline {N}}^c_{\overline H}>=M_{\rm H}$. In addition, the
superfields $H$ and ${\overline H}$ are eaten and acquire large masses via
the supersymmetric Higgs mechanism, except for $D_H^c$ and 
${\overline {D}}^c_{\overline H}$. And the superpotential $ \lambda_1 H H h$ and
$ \lambda_2 {\overline H} {\overline H} {\overline h}$ couple the $D_H^c$ and
${\overline {D}}^c_{\overline H}$ with the $D_h$ and ${\overline {D}}_{\overline h}$,
respectively, to form the massive eigenstates with masses
$2 \lambda_1 <N_H^c>$ and $2 \lambda_2 <{\overline {N}}^c_{\overline H}>$. So, we
naturally have the doublet-triplet splitting due to the missing
partner mechanism~\cite{AEHN-0}. 
Because the triplets in $h$ and ${\overline h}$ only have
small mixing through the $\mu$ term, the Higgsino-exchange mediated
proton decay are negligible, {\it i.e.},
we do not have the dimension-5 proton
decay problem.

The SM fermion masses are from the following
superpotential
\bea 
{ W}_{\rm Yukawa} = y_{ij}^{D}
F_i F_j h + y_{ij}^{U \nu} F_i  {\overline f}_j {\overline
h}+ y_{ij}^{E} {\overline l}_i  {\overline f}_j h + \mu h {\overline h}
+ y_{ij}^{N} \phi_i {\overline H} F_j~,~\,
\label{potgut}
\eea
where $y_{ij}^{D}$, $y_{ij}^{U \nu}$, $y_{ij}^{E}$ and $y_{ij}^{N}$
are Yukawa couplings, and $\mu$ is the bilinear Higgs mass term.

After the $SU(5)\times U(1)_X$ gauge symmetry is broken down to the SM gauge 
symmetry, the above superpotential gives 
\bea 
{ W_{SSM}}&=&
y_{ij}^{D} D^c_i Q_j H_d+ y_{ji}^{U \nu} U^c_i Q_j H_u
+ y_{ij}^{E} E^c_i L_j H_d+  y_{ij}^{U \nu} N^c_i L_j H_u \nnb \\
&& +  \mu H_d H_u+ y_{ij}^{N} 
\langle {\overline {N}}^c_{\overline H} \rangle \phi_i N^c_j
 + \cdots (\textrm{decoupled below $M_{GUT}$}). 
\label{poten1}
\eea

\subsection{Higgs Boson Mass}

Since we will consider the  high-scale supersymmetry 
and split supersymmetry,  let us briefly review
how to calculate the Higgs boson masses in these scenarios.

We denote the gauge couplings for
the $U(1)_Y$, $SU(2)_L$, and $SU(3)_C$ as
 $g_Y$, $g_2$, and $g_3$, respectively,
and define $g_1\equiv {\sqrt {5/3}} g_Y$.
The major prediction in the models with high-scale supersymmetry
or split supersymmetry is the Higgs boson
 mass~\cite{Barger:2004sf, Barger:2005gn, NASD, Giudice:2004tc}.
 We can calculate the Higgs boson quartic coupling $\lambda$ at
the supersymmetry breaking scale $M_S$~\cite{Barger:2004sf, NASD}
\begin{equation}
\lambda({M_S}) = \frac{ g_2^2(M_S) + 3 g_1^2(M_S)/5}{4} \cos^2 2\beta~,
\end{equation}
where $\beta$ is the mixing angle of one pair of Higgs doublets
in the supersymmetric Standard 
Models~\cite{Barger:2004sf, Barger:2005gn, NASD, Giudice:2004tc}.
And then we evolve it down to the weak scale. The renormalization group
equation for the Higgs quartic coupling is given in Appendix~\ref{apdxA}.
Using the one-loop effective Higgs potential with top quark radiative
corrections, we calculate the Higgs boson mass by minimizing the
one-loop effective potential
\be
V_{eff} = m_h^2 H^\dagger H - \frac{\lambda}{2!} (H^\dagger H)^2 -
\frac{3}{16\pi^2} h_t^4 (H^\dagger H)^2 \left[\log\frac{h_t^2 (H^\dagger
H)}{Q^2} - \frac{3}{2}\right]\,,
\ee
where $m_h^2$ is the bare Higgs mass square,
$h_t$ is the top quark Yukawa coupling, and the scale $Q$ is
chosen to be at the Higgs boson mass. 
For the ${\overline{MS}}$ top quark Yukawa coupling, we use the one-loop
corrected value~\cite{Arason:1991ic}, which is related to the top quark pole
mass by
\be
m_t = h_t v \left(1 + \frac{16}{3}\frac{g_3^2}{16\pi^2} - 2\frac{h_t^2}{16\pi^2}\right)\,.
\ee

We define $\alpha_i=g_i^2/4\pi$ and denote the $Z$ boson mass as $M_Z$.
In the following numerical calculations, we vary $\tan\beta$ from
1.5 to 50. We choose the recent top quark pole mass measurement
$m_t = 172.7 \pm 2.9~\GeV$~\cite{:2005cc}, the strong coupling constant
$\alpha_3(M_Z) = 0.1182 \pm 0.0027$~\cite{Bethke:2004uy}, and the
fine structure constant $\alpha_{EM}$, weak mixing angle $\theta_W$ and 
Higgs vacuum expectation value (VEV) $v$ at $M_Z$ to be~\cite{Eidelman:2004wy}
\bea
&&\alpha^{-1}_{EM}(M_Z) = 128.91 \pm 0.02\,, \nonumber \\
&&\sin^2\theta_W(M_Z) = 0.23120 \pm 0.00015\,, \nonumber \\
&&v = 174.10\,{\rm GeV}\,.
\eea

\section{String-Scale Gauge Coupling Unification}

To achieve string-scale gauge coupling unification, we introduce 
 vector-like particles which form complete 
flipped $SU(5)\times U(1)_X$ multiplets.
The quantum numbers for these additional vector-like particles
 under the $SU(5)\times U(1)_X$ gauge symmetry are
\begin{eqnarray}
&& XF ={\mathbf{(10, 1)}}~,~{\overline{XF}}={\mathbf{({\overline{10}}, -1)}}~,~\\
&& Xf={\mathbf{(5, 3)}}~,~{\overline{Xf}}={\mathbf{({\overline{5}}, -3)}}~,~\\
&& Xl={\mathbf{(1, -5)}}~,~{\overline{Xl}}={\mathbf{(1, 5)}}~,~\\
&& Xh={\mathbf{(5, -2)}}~,~{\overline{Xh}}={\mathbf{({\overline{5}}, 2)}}~.~\,
\end{eqnarray}
It is obvious that $XF$, ${\overline{XF}}$, $Xf$, ${\overline{Xf}}$,
$Xl$, ${\overline{Xl}}$, $Xh$, and ${\overline{Xh}}$ are standard vector-like 
particles.

Moreover,  the particle contents for
$XF$, ${\overline{XF}}$, $Xf$, ${\overline{Xf}}$,
$Xl$, ${\overline{Xl}}$, $Xh$, and ${\overline{Xh}}$ are
\begin{eqnarray}
&& XF = (XQ, XD^c, XN^c)~,~ {\overline{XF}}=(XQ^c, XD, XN)~,~\\
&& Xf=(XU, XL^c)~,~ {\overline{Xf}}= (XU^c, XL)~,~\\
&& Xl= XE~,~ {\overline{Xl}}= XE^c~,~\\
&& Xh=(XD, XL)~,~ {\overline{Xf}}= (XD^c, XL^c)~.~
\end{eqnarray}
Under the $SU(3)_C \times SU(2)_L \times U(1)_Y$ gauge
symmetry, the quantum numbers for the extra vector-like 
particles are
\begin{eqnarray}
&& XQ={\mathbf{(3, 2, {1\over 6})}}~,~
XQ^c={\mathbf{({\bar 3}, 2,-{1\over 6})}} ~,~\\
&& XU={\mathbf{({3},1, {2\over 3})}}~,~
XU^c={\mathbf{({\bar 3},  1, -{2\over 3})}}~,~\\
&& XD={\mathbf{({3},1, -{1\over 3})}}~,~
XD^c={\mathbf{({\bar 3},  1, {1\over 3})}}~,~\\
&& XL={\mathbf{({1},  2,-{1\over 2})}}~,~
XL^c={\mathbf{(1,  2, {1\over 2})}}~,~\\
&& XE={\mathbf{({1},  1, {-1})}}~,~
XE^c={\mathbf{({1},  1, {1})}}~,~\\
&& XN={\mathbf{({1},  1, {0})}}~,~
XN^c={\mathbf{({1},  1, {0})}}~,~\\
&& XY={\mathbf{({3}, 2, -{5\over 6})}}~,~
XY^c={\mathbf{({\bar 3}, 2, {5\over 6})}} ~.~\
\end{eqnarray}

To have the string-scale gauge coupling unification and avoid the
Landau pole problem, we need to introduce sets of 
vector-like particles at the TeV scale whose contributions to the one-loop
beta functions satisfy $\Delta b_1 < \Delta b_2 = \Delta b_3$.
To avoid the Landau pole problem, we find that there are only four
 possible such sets of vector-like
 particles as follows
due to the quantizations of the one-loop beta functions
\begin{eqnarray}
&& Z0: XF+{\overline{XF}}~;~\\
&& Z1: XF+{\overline{XF}}+Xf+{\overline{Xf}}~;~\\
&& Z2: XF+{\overline{XF}}+Xl+{\overline{Xl}}~;~\\
&& Z3: XF+{\overline{XF}}+Xl+{\overline{Xl}}
+Xh+{\overline{Xh}}~.~\,
\end{eqnarray}
We assume the masses for each set of vector-like particles are the same,
and denote them as $M_V$. In the interesting flipped $SU(5)\times U(1)_X$
 string models from the four-dimensional fermionic formulation of the 
weakly coupled heterotic string theory~\cite{LNY}, for example, the
so called 5/2 model in Table 4 in Ref.~\cite{LNY}, in addition to the 
particle content in the flipped $SU(5)\times U(1)_X$ model reviewed 
in the subsection IIA, we have one pair of vector-like particles 
$XF$ and ${\overline{XF}}$, one pair $Xf$ and ${\overline{Xf}}$,
one pair $Xl$ and ${\overline{Xl}}$, and two pairs
$Xh$ and ${\overline{Xh}}$.  Moreover, some of
these vector-like particles can be light and have masses around
the TeV scale, while the others may have masses around the string
scale. So, we can indeed obtain the above $Zi$ sets of vector-like
particles with masses around the TeV scale.

In addition, to obtain the exact string-scale gauge coupling unification,
we may need to consider the string-scale (or GUT-scale) threshold
corrections. Interestingly, in the same 5/2 model in Table 4 
in Ref.~\cite{LNY}, there exist additional gauge symmetries
$SO(10)\times SU(4)\times U(1)^5$ in the hidden sector.
Also, there are five pairs of vector-like particles $XT_i$
and ${\overline{XT}}_i$ and one pair of vector-like particles ${ XT'}$
and ${\overline{XT'}}$~\cite{LNY}. These particles are only charged
under $SU(4) \times U(1)_X$, and their quantum numbers are
\begin{eqnarray}
&&  XT_i={\mathbf{(4,  {5\over 2})}}~,~
\overline{XT}_i={\mathbf{({\bar 4}, -{5\over 2})}} ~,~ \\
&& XT'={\mathbf{(4,  -{5\over 2})}}~,~
\overline{XT'}={\mathbf{({\bar 4}, {5\over 2})}} ~.~\,
\end{eqnarray}
To mimic the string-scale threshold corrections, 
we introduce this set of vector-like particles
 with masses  close to the string scale.
We denote this set of vector-like particles as
ZT set. For simplicity, we assume that
the masses for ZT set of vector-like particles
are universal, and we denote this mass
as $M_{V'}$.


In this paper, we assume that $H$, ${\overline{H}}$, and the
triplets $D_h$ and ${\overline {D}}_{\overline h}$
in $h$ and ${\overline h}$ have masses around the $ SU(2)_L \times SU(3)_C$
unification scale $M_{23}$. From the weak scale to $M_{23}$,
we employ two-loop RGE running for the SM
gauge couplings and one-loop running for the Yukawa couplings.
 For simplicity, we only consider the
contributions to the gauge coupling RGE running from the Yukawa
couplings of the third family of the SM fermions, \ie, the top quark,
bottom quark and $\tau$ lepton Yukawa couplings. And we neglect
the contributions to the gauge coupling RGE running from the Yukawa
couplings of the extra vector-like particles, and
the threshold corrections at the supersymmetry breaking scale
due to the scalar mass differences.

From $M_{23}$ to $M_{\rm string}$, we consider two-loop RGE running for the 
$SU(5)\times U(1)_X$ gauge couplings. For simplicity, we neglect
the Yukawa coupling corrections.
The RGEs and beta functions are given in the Appendicies A and B.
Also, the beta functions for $XT'$ and $\overline{XT'}$ are the same as 
these for $XT_i$ and $\overline{XT}_i$, and then will not be
 presented there.

In addition, we would like to point out that 
the gauge coupling  $\alpha_1'$ of $U(1)_X$ is related to
$\alpha_1$ and $\alpha_5$ at the scale $M_{23}$ by
\bea
\alpha_1'^{-1}(M_{23}) = \frac{25}{24} \alpha_1^{-1}(M_{23}) -
\frac{1}{24} \alpha_5^{-1}(M_{23})~.
\eea



\subsection{Universal Supersymmetry Breaking}

We consider the following cases for the vector-like particle mass scales $M_V$
and supersymmetry breaking scales $M_S$

(1) $M_V=200$ GeV, $M_S=360$ GeV, $1000$ GeV, and $1.0\times 10^4$ GeV;

(2) $M_V = 600$ GeV, $M_S=1000$ GeV;

(3)  $M_V=1000$ GeV, $M_S=1.0\times 10^4$ GeV.

In our numerical calculations, 
the RGE running is performed for $\tan\beta = 10$.
Varying its value will only generate negligible changes to the mass scales.
However, the choice of $\tan\beta$ affects the Higgs boson mass range
significantly if the supersymmetry breaking scale is high. 
 The Higgs boson mass ranges are shown with the lower
end calculated with $\alpha_3 = 0.1209$, $m_t = 169.8$ GeV and $\tan\beta
= 1.5$, and the upper end with $\alpha_3 = 0.1155$, $m_t = 175.6$ GeV and
$\tan\beta = 50$.  

\begin{table}[htb]
\begin{center}
\begin{tabular}{|c|c|c|c|c|c|c|c|}
\hline
 &  $M_V$  &  $M_S$  & $M_{23}$  &  $M_{V'}$ & $g_{\rm
string}$ & $M_{\rm string}$ & $m_h$ \\
\hline
Z2 & 200   & 360 & $1.9 \times 10^{16}$  & $1.1 \times
10^{17}$ & 1.301 & $6.9 \times 10^{17}$ &  $-$  \\
Z2 & 200   & 1000 & $1.9 \times 10^{16}$ & $1.5 \times 10^{17}$ &
1.207 & $6.4 \times 10^{17}$ &  $-$  \\
Z2 & 200   & $1.0 \times 10^4$ & $2.1 \times 10^{16}$  & $2.7 \times
10^{17}$ & 1.064 & $5.6 \times 10^{17}$ & 102   $-$ 132 \\
Z2 & 600   & 1000 & $1.8 \times 10^{16}$ & $1.8 \times 10^{17}$ &
1.170 & $6.2 \times 10^{17}$ &  $-$   \\
Z2 & 1000  & $1.0 \times 10^4$ & $2.0 \times 10^{16}$  & $3.5 \times
10^{17}$ & 1.031 & $5.4 \times 10^{17}$ & 102  $-$ 132 \\
\hline
\end{tabular}
\end{center}
\caption{Mass scales in the flipped $SU(5)\times U(1)_X$ models
with string-scale gauge coupling unification and universal
supersymmetry breaking.}
\label{tbl:univ}
\end{table}

\begin{figure}[htb]
\centering
\includegraphics[width=10cm]{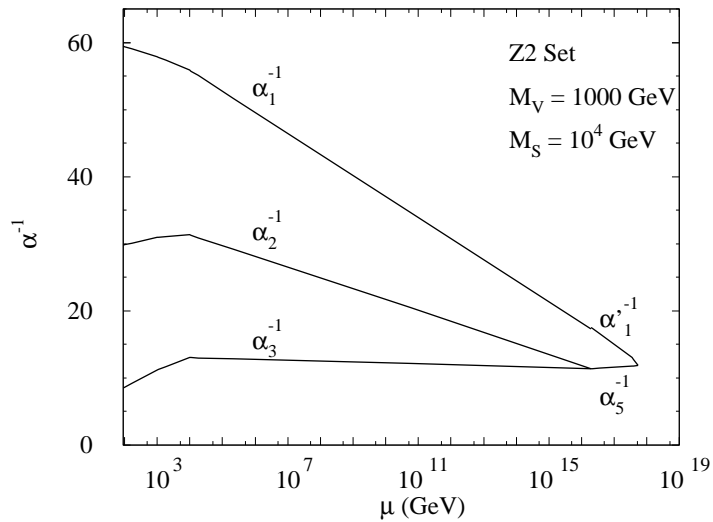}
\caption{Two-loop string-scale gauge coupling unification for
universal supersymmetry breaking in the
flipped $SU(5)\times U(1)_X$ model with Z2 set of vector-like
particles, $M_V = 1000$ GeV, and $M_S = 1.0\times 10^4$ GeV.} 
\label{fig:z2cbx}
\end{figure}

From the concrete numerical calculations, 
we obtain that the string-scale gauge coupling
unification can not be achieved in all the cases with 
Z1 and Z3 sets of vector-like particles,
because the strong coupling $g_3$ runs into Landau
pole below the $SU(2)_L\times SU(3)_C$
unification scale $M_{23}$. The string-scale
gauge coupling unification can be realized precisely in
the cases with Z2 set of vector-like particles and 
the suitable mass $M_{V'}$ for the
ZT set of vector-like particles. In the Table~\ref{tbl:univ},
we present the corresponding $M_{23}$, $M_{V'}$, string
coupling $g_{\rm string}$, string-scale $M_{\rm string}$,
and the Higgs boson mass $m_h$ if the supersymmetry breaking 
scale is high. We find that $M_{V'}$ is close to
the string scale. Thus, the vector-like particles 
 ($XT_i$, $\overline{XT}_i$), and 
($XT'$, $\overline{XT'}$) can indeed be considered as
the string-scale threshold corrections.
Moreover, if the supersymmetry breaking scale is
about $1.0\times 10^{4}$ GeV, the corresponding Higgs boson masses are 
from $102$ GeV to $132$ GeV. As an example, we present the
two-loop string-scale gauge coupling unification in the model 
with Z2 set of vector-like
particles, $M_V = 1000$ GeV, and $M_S = 1.0\times 10^4$ GeV
in Fig.~\ref{fig:z2cbx}.

\subsection{Split Supersymmetry}

The RGEs and beta functions in the split supersymmetry can be found in
Ref.~\cite{Giudice:2004tc}. For simplicity, we assume that the
gaugino and Higgsino masses are the same and equal to the 
 vector-like particle mass scale $M_{V}$.
We consider the following cases for  $M_V$
and supersymmetry breaking scale $M_S$ that is the universal scalar mass
in split supersymmetry

(1) $M_V=200$ GeV, $M_S=1.0\times 10^4$ GeV, and $1.0\times 10^{10}$ GeV;

(2)  $M_V=1000$ GeV, $M_S=1.0\times 10^4$ GeV, and $1.0\times 10^{10}$ GeV.

\begin{table}[htb]
\begin{center}\begin{tabular}{|c|c|c|c|c|c|c|c|}
\hline
 &  $M_V$  &  $M_S$  & $M_{23}$  &  $M_{V'}$ & $g_{\rm
string}$ & $M_{\rm string}$ & $m_h$ \\
\hline
Z1 & 200   & $1.0 \times 10^{10}$  & $1.2 \times 10^{17}$ & $3.9 \times
10^{16}$ & 1.179 & $6.2 \times 10^{17}$  &  122 $-$ 149 \\
Z1 & 1000  & $1.0 \times 10^{10}$ & $8.7 \times 10^{16}$ & $1.6 \times
10^{17}$ & 1.005 & $5.3 \times 10^{17}$ & 126 $-$ 151 \\
\hline
Z2 & 200   & $1.0 \times 10^4$  & $2.6 \times 10^{16}$  & $9.2 \times
10^{16}$ & 1.171 & $6.2 \times 10^{17}$ &   104 $-$ 134 \\
Z2 & 200   & $1.0 \times 10^{10}$  & $7.9 \times 10^{16}$ & $3.6 \times
10^{16}$ & 0.918 & $4.8 \times 10^{17}$  &  123 $-$ 149 \\
Z2 & 1000  & $1.0 \times 10^4$  & $2.2 \times 10^{16}$  & $1.9 \times
10^{17}$ & 1.081 & $5.7 \times 10^{17}$ &   104 $-$ 133 \\
Z2 & 1000  & $1.0 \times 10^{10}$ & $6.2 \times 10^{16}$ & $1.5 \times
10^{17}$ & 0.842 & $4.4 \times 10^{17}$  &  127 $-$ 151 \\
\hline
Z3 & 200   & $1.0 \times 10^{10}$  & $1.3 \times 10^{17}$ & $3.1 \times
10^{16}$ & 1.160 & $6.1 \times 10^{17}$  &   122 $-$ 149 \\
Z3 & 1000  & $1.0 \times 10^{10}$ & $8.6 \times 10^{16}$ & $1.4 \times
10^{17}$ & 1.005 & $5.3 \times 10^{17}$ & 126 $-$ 151 \\
\hline
\end{tabular}
\end{center}
\caption{Mass scales in the flipped $SU(5)\times U(1)_X$ models
with string-scale gauge coupling unification and split
supersymmetry.}
\label{tbl:split}
\end{table}

\begin{figure}[htb]
\centering
\includegraphics[width=10cm]{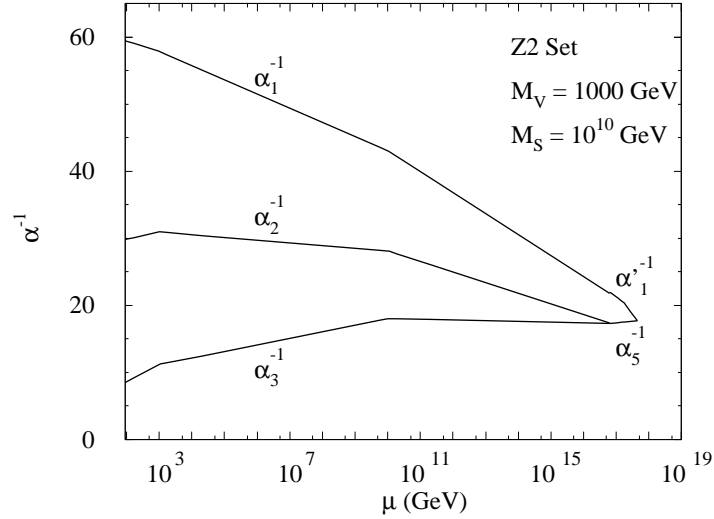}
\caption{Two-loop string-scale gauge coupling unification for
split supersymmetry in the
flipped $SU(5)\times U(1)_X$ model with Z2 set of vector-like
particles, $M_V = 1000$ GeV, and $M_S = 10^{10}$ GeV.} 
\label{fig:z2cdx}
\end{figure}

If supersymmetry breaking scale is $1.0\times 10^{4}$ GeV,
we find that the string-scale gauge coupling
unification can not be achieved in all the cases with 
Z1 and Z3 sets of vector-like particles,
because there exists the Landau pole problem for 
the strong coupling $g_3$ below $M_{23}$. The string-scale
gauge coupling unification can be realized precisely in
 the rest cases with suitable
$M_{V'}$. In Table~\ref{tbl:split},
we present the corresponding $M_{23}$, $M_{V'}$, 
 $g_{\rm string}$, $M_{\rm string}$,
and $m_h$. Because $M_{V'}$ is also close to
the string scale (all within one order), the vector-like particles
 ($XT_i$, $\overline{XT}_i$), and
($XT'$, $\overline{XT'}$) can be considered as
 string-scale threshold corrections, too.
Moreover, if the supersymmetry breaking scale is
about $1.0\times 10^{4}$ GeV, the corresponding 
 Higgs boson masses range from
$102$ GeV to $134$ GeV. And if the supersymmetry breaking
scale is $1.0\times 10^{10}$ GeV, the Higgs boson masses
are from $122$ GeV to $151$ GeV. In Fig.~\ref{fig:z2cdx},
we present the two-loop string-scale gauge coupling unification in the
 model with Z2 set of vector-like particles,
 $M_V = 1000$ GeV, and $M_S = 10^{10}$ GeV.

\section{The Simple Flipped $SU(5)\times U(1)_X$ Models}

First, we consider a simple flipped $SU(5)\times U(1)_X$ model
 with high-scale supersymmetry breaking. Interestingly, we can achieve 
the string-scale gauge coupling unification by  introducing
only one Z0 set of vector-like particles around the TeV scale and
 choosing suitable supersymmetry breaking scale.

\begin{table}[htb]
\begin{center}
\begin{tabular}{|c|c|c|c|c|c|c|}
\hline
 &  $M_V$  &  $M_S$  & $M_{23}$  &  $g_{\rm
string}$ & $M_{\rm string}$ & $m_h$ \\
\hline
Z0 & 200 & $5.8 \times 10^{12}$ & $4.4 \times 10^{16}$ & 0.664 & $3.5
\times 10^{17}$ & 128 - 142 \\ 
Z0 & 1000 & $1.4 \times 10^{12}$ & $3.9 \times 10^{16}$ & 0.670 & $3.5
\times 10^{17}$ & 131 - 144 \\ 
\hline
\end{tabular}
\end{center}
\caption{Mass scales in the simple flipped $SU(5)\times U(1)_X$ model
with string-scale gauge coupling unification and
intermediate-scale supersymmetry breaking.}
\label{tbl:Z0}
\end{table}

\begin{figure}[htb]
\centering
\includegraphics[width=10cm]{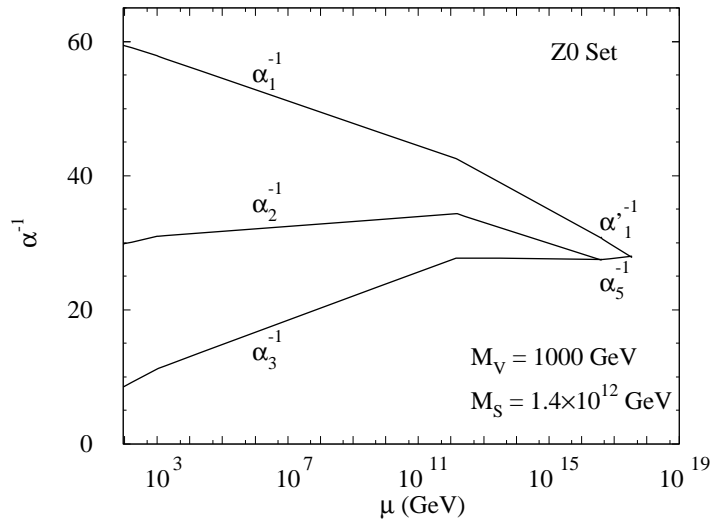}
\caption{Two-loop string-scale gauge coupling unification
for the simple flipped $SU(5)\times U(1)_X$ model
 with $M_V = 1000$ GeV and intermediate-scale  
supersymmetry breaking.} 
\label{fig:z0}
\end{figure}

For numerical calculations, we choose $M_V=200$ GeV and
$1000$ GeV. We obtain that the supersymmetry breaking
scale is about $10^{12}$ GeV, the $SU(2)_L\times SU(3)_C$
unification scale is about $4\times 10^{16}$ GeV, and
the string scale is about $3.5\times 10^{17}$ GeV.
The concrete results including the Higgs boson masses
are given in Table~\ref{tbl:Z0}. For $M_V=1000$ GeV,
we present the two-loop string-scale gauge coupling unification in
Fig~\ref{fig:z0}.

As we know, the Peccei-Quinn mechanism provides an
elegant solution to the strong CP problem~\cite{PQ}. However,
the Peccei-Quinn mechanism may not be stable against
the quantum gravity corrections.
And the Peccei-Quinn mechanism may be 
 stabilized if and only if the supersymmetry breaking
scale is around $10^{11-12}$ GeV~\cite{Babu:2002ic, Barger:2004sf}. 
Interestingly, our supersymmetry breaking scales are  within
this range.

\begin{table}[htb]
\begin{center}
\begin{tabular}{|c|c|c|c|c|c|}
\hline
 &  $M_S$  &  $M_V$  & $M_{23}$  &  $g_{\rm string}$ & $M_{\rm string}$  \\
\hline
Z0 & 200  & $4.5 \times 10^{11}$ & $1.6 \times 10^{16}$ & 0.800 & $4.2
\times 10^{17}$ \\ 
Z0 & 1000 & $1.6 \times 10^{11}$ & $1.6 \times 10^{16}$ & 0.789 & $4.2
\times 10^{17}$  \\ 
\hline
\end{tabular}
\end{center}
\caption{Mass scales in the simple flipped $SU(5)\times U(1)_X$ model
with string-scale gauge coupling unification and
intermediate-scale mass for the
vector-like particles.}
\label{tbl:Z0inv}
\end{table}

\begin{figure}[htb]
\centering
\includegraphics[width=10cm]{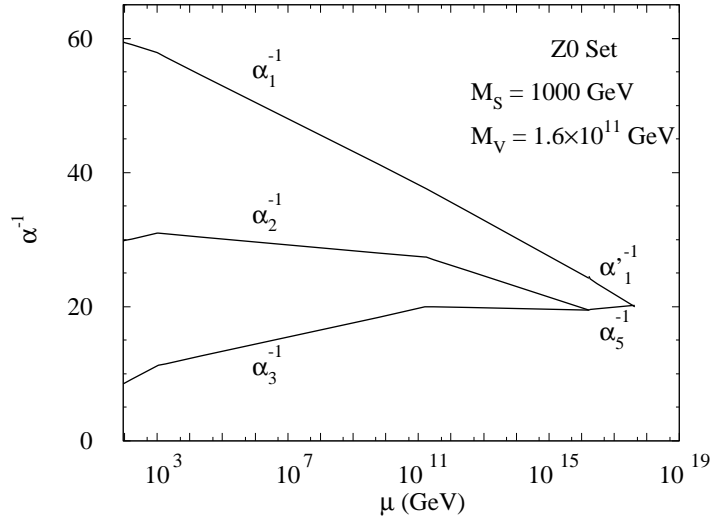}
\caption{Two-loop string-scale gauge coupling unification
for the simple flipped $SU(5)\times U(1)_X$ model
 with intermediate-scale  universal mass for the
vector-like particles.}
\label{fig:z0inv}
\end{figure}

Second, the string-scale gauge coupling unification 
can also be achieved in the flipped $SU(5)\times U(1)_X$
models with low-energy supersymmetry 
by introducing only one Z0 set of vector-like particles 
at the intermediate scale, which has been
studied previously~\cite{Lopez:1993qn, Lopez:1995cs}. 
Using the updated
gauge couplings at the weak scale, we
are able to reproduce the string-scale 
gauge coupling unification in this case, too.
With $M_S=200$ GeV and $1000$ GeV, we
present the $M_V$, $M_{23}$,
$g_{\rm string}$  and $M_{\rm string}$
in Table~\ref{tbl:Z0inv}.
We find that the masses for 
the vector-like particles are about
$10^{11}$ GeV, which are about two
orders higher than the previous 
results~\cite{Lopez:1993qn, Lopez:1995cs}. For $M_S=1000$ GeV,
we present the two-loop string-scale gauge coupling unification in
Fig~\ref{fig:z0inv}.

\section{Comments}
In our models, the $XN^c$ in $XF$ and $XN$ in $\overline{XF}$ can 
have masses around the scale $M_{23}$, and the SM fermions may 
mix with the additional vector-like particles. For example, we
consider the model with $Z_2$ set of vector-like particles.
The relevant Langrangian is 
\bea 
{ W}_{\rm Yukawa} &=& y_{ij}^{D}
F_i F_j h + y_{ij}^{U \nu} F_i  {\overline f}_j {\overline
h}+ y_{ij}^{E} {\overline l}_i  {\overline f}_j h + \mu h {\overline h}
+ y_{kj}^{N} \phi_k {\overline H} F_j + y_{j}^{\prime D} XF  F_j h \nnb \\
&& 
+ y_{j}^{\prime U \nu} XF  {\overline f}_j {\overline
h}+ y_{j}^{\prime E} \overline{Xl}  {\overline f}_j h 
+ y_{k}^{\prime N} \phi_{k} {\overline H} XF 
+ y^{\overline{XF}}  \phi^{\prime} H \overline{XF}
~,~\,
\label{potgut-Comments}
\eea
where $k=1, 2, 3, 4$, and
 $\phi_4$ and $\phi^{\prime}$ are additional SM singlets.
Therefore,  the $XN^c$ in $XF$ and $XN$ in $\overline{XF}$ obtain
 masses around the scale $M_{23}$. And the SM fermions (including
neutrinos) will mix
with the corresponding vector-like particles.
In particular, the additional vector-like partilces,
for example, $XF$ and $\overline{XF}$, and/or $Xf$ and
$\overline{Xf}$, will definitely affect the discussions
of neutrino masses due to the extra mixing terms. The concerete 
discussions of the mixings among the  SM fermions and the 
vector-like particles
are beyond the scope of this paper.

\section{Discussion and Conclusions}

Whether we can test GUTs in the future experiments is an interesting
question. We pointed out that the little hierarchy between
the GUT scale and the string scale may be tested at the LHC. 
To realize the precise string-scale gauge coupling unification, 
we introduce the
additional vector-like particles. We require that these vector-like
particles be standard, form complete GUT multiplets, and
have masses around the TeV scale or close to the string scale.
The vector-like particles with TeV-scale masses can be observed at the
LHC, and the vector-like particles with masses close to the string-scale
can be considered as the string-scale threshold corrections.
We found that only the flipped $SU(5)\times U(1)_X$ models 
can work elegantly. Moreover, we listed all the possible sets of 
vector-like particles with masses around the TeV scale. 
And we introduce vector-like
particles with masses close to the string scale which can mimic
the string-scale threshold corrections. We emphasize that
all of these vector-like particles can be obtained in the 
interesting flipped $SU(5)\times U(1)_X$ string models from the
four-dimensional free fermionic formulation of
the weakly coupled heterotic string theory~\cite{LNY}.
 Assuming the low-energy supersymmetry, high-scale supersymmetry,
or  split supersymmetry, we show that the string-scale
gauge coupling unification can indeed be achieved in the
flipped $SU(5)\times U(1)_X$ models. These models can
be tested at the LHC  by observing the simple sets of
vector-like particles with masses around the TeV scale. In addition,
we  discuss a simple flipped $SU(5)\times U(1)_X$ model 
with string-scale gauge coupling unification and
high-scale supersymmetry by introducing only
one pair of the vector-like particles at the
TeV scale, and we predict the corresponding Higgs boson masses.
Also, we briefly comment on the string-scale gauge coupling
unification in the model with low-energy supersymmetry 
and only one pair of vector-like
particles with masses at the intermediate scale.
And we briefly comment on the mixings among the SM fermions 
and the corresponding extra
vector-like particles.

\begin{acknowledgments}

JJ thanks the National Center for Theoretical Sciences 
in Taiwan for its hospitality, where part of the work was done.
This research was supported by the U.S.~Department of Energy
under Grants No. (JJ) DE-FG02-96ER40969, (TL) DE-FG02-96ER40959,
and (DVN) DE-FG03-95-Er-40917, and by the Cambridge-Mitchell Collaboration 
in Theoretical Cosmology (TL).

\end{acknowledgments}

\appendix

\section{ The Non-Supersymmetric and Supersymmetric Standard
Models with  Vector-Like Particles}
\label{apdxA}

\subsection{ Renormalization Group Equations}

We give the renormalization group equations
in the SM and MSSM.
The general formulae for the renormalization group equations
in the SM are given in Refs.~\cite{mac,Cvetic:1998uw}, and 
these for the supersymmetric models
are given in Refs.~\cite{Barger:1992ac,Barger:1993gh,Martin:1993zk}.

First, we summarize the  renormalization group equations
in the SM.
The two-loop renormalization group equations for the gauge couplings are
\begin{eqnarray}
(4\pi)^2\frac{d}{dt}~ g_i=b_i g_i^3 &+&\frac{g_i^3}{(4\pi)^2}
\left[ \sum_{j=1}^3B_{ij}g_j^2-\sum_{\alpha=u,d,e}d_i^\alpha
{\rm Tr}\left( h^{\alpha \dagger}h^{\alpha}\right) \right] ~,~\,
\label{SMgauge}
\end{eqnarray}
where $t=\ln  \mu$ and $ \mu$ is the renormalization scale.
The $g_1$, $g_2$ and $g_3$ are the gauge couplings
for $U(1)_Y$, $SU(2)_L$ and $SU(3)_C$, respectively,
where we use the $SU(5)$ normalization $g_1^2 \equiv (5/3)g_Y^{ 2}$.
The beta-function coefficients are  
\begin{eqnarray}
&&b=\left(\frac{41}{10},-\frac{19}{6},-7\right) ~,~
B=\pmatrix{\frac{199}{50}&
\frac{27}{10}&\frac{44}{5}\cr \frac{9}{10} & \frac{35}{6}&12 \cr
\frac{11}{10}&\frac{9}{2}&-26} ~,~\\
&&d^u=\left(\frac{17}{10},\frac{3}{2},2\right) ~,~
d^d=\left(\frac{1}{2},\frac{3}{2},2\right) ~,~
d^e=\left(\frac{3}{2},\frac{1}{2},0\right) ~.~\,
\end{eqnarray}

Since the contributions in Eq.~(\ref{SMgauge}) from
the Yukawa couplings arise from  the 
two-loop diagrams, we only need Yukawa coupling
 evolution at the one-loop order.
The one-loop renormalization group equations for Yukawa couplings are
\begin{eqnarray}
(4\pi)^2\frac{d}{dt}~h^u&=&h^u\left( -\sum_{i=1}^3c_i^ug_i^2
+\frac{3}{2}
h^{u \dagger}h^{u}
-\frac{3}{2}
h^{d \dagger}h^{d}
+\Delta_2\right) ~,~\\
(4\pi)^2\frac{d}{dt}~h^d&=&h^d\left( -\sum_{i=1}^3c_i^dg_i^2
-\frac{3}{2}
h^{u \dagger}h^{u}
+\frac{3}{2}
h^{d \dagger}h^{d}
+\Delta_2 \right)~,~\\
(4\pi)^2\frac{d}{dt}~h^e&=&h^e\left( -\sum_{i=1}^3c_i^eg_i^2
+\frac{3}{2}
h^{e \dagger}h^{e}
+ \Delta_2 \right) ~,~\,
\label{SMY}
\end{eqnarray}
where $h^u$, $h^d$ and $h^e$ are the Yukawa couplings
for the up-type quark, down-type quark, and lepton,
respectively. Also, $c^u$, $c^d$, and $c^e$ are given by 
\begin{eqnarray}
c^u=\left( \frac{17}{20}, \frac{9}{4}, 8\right) ~,~
c^d=\left( \frac{1}{4}, \frac{9}{4}, 8\right) ~,~
c^e=\left( \frac{9}{4}, \frac{9}{4}, 0\right) ~,~
\end{eqnarray}
\begin{eqnarray}
\Delta_2 &=& {\rm Tr} ( 3h^{u \dagger}h^{u}+3 h^{d \dagger}h^{d}+
h^{e \dagger}h^{e})  ~.~
\end{eqnarray}

The one-loop renormalization group equation for the Higgs quartic coupling is 
\begin{eqnarray}
(4\pi)^2\frac{d}{dt}~\lambda  &=&12 \lambda^2
-\left({9\over 5} g_1^2 + 9 g_2^2 \right) \lambda
+{9\over 4} \left( {3\over {25}} g_1^4 
+ {2\over 5} g_1^2g_2^2 + g_2^4 \right)
+4\Delta_2 \lambda - 4 \Delta_4 ~,~\,
\end{eqnarray}
where
\begin{eqnarray}
\Delta_4 &=& {\rm Tr} \left[ 3 (h^{u \dagger}h^{u})^2+3 (h^{d \dagger}h^{d})^2
+ (h^{e \dagger}h^{e})^2\right]  ~.~
\end{eqnarray}

Second, we summarize the renormalization group equations in 
the MSSM.
The two-loop renormalization group equations for the gauge couplings are
\begin{eqnarray}
(4\pi)^2\frac{d}{dt}~ g_i &=& b_i g_i^3 
 +\frac{g_i^3}{(4\pi)^2}
\left[~ \sum_{j=1}^3 B_{ij}  g_j^2-\sum_{\alpha=u,d,e}d_i^\alpha
{\rm Tr}\left( y^{\alpha \dagger}y^{\alpha}\right) \right] ~,~\,
\label{SUSYgauge}
\end{eqnarray}
where the beta-function coefficients are 
\begin{eqnarray}
&&b=\left(\frac{33}{5},1,-3\right) ~,~~~  B=\pmatrix{\frac{199}{25}&
\frac{27}{5}&\frac{88}{5}\cr \frac{9}{5} & 25&24 \cr
\frac{11}{5}&9&14} ~,~\\
&&d^u=\left(\frac{26}{5},6,4\right) ~,~
d^d=\left(\frac{14}{5},6,4\right) ~,~
d^e=\left(\frac{18}{5},2,0\right) ~.~ \\
\end{eqnarray}

The one-loop renormalization group equations for Yukawa couplings are
\begin{eqnarray}
(4\pi)^2\frac{d}{dt}~y^u&=& y^u
\left[ 3 y^{u \dagger} y^{u}+ y^{d \dagger} y^{d}
+3{\rm Tr}( y^{u \dagger} y^{u}) 
-\sum_{i=1}^3c_i^ug_i^2 \right]~,~\\
(4\pi)^2\frac{d}{dt}~y^d&=& y^d
\left[ y^{u \dagger} y^{u} + 3 y^{d \dagger} y^{d}
+{\rm Tr}(3 y^{d \dagger} y^{d}
+ y^{e \dagger} y^{e}) 
-\sum_{i=1}^3c_i^dg_i^2 \right]~,~\\
(4\pi)^2\frac{d}{dt}~y^e&=& y^e
\left[ 3 y^{e \dagger} y^{e}+{\rm Tr}(3 y^{d \dagger} y^{d}
+ y^{e \dagger} y^{e}) 
-\sum_{i=1}^3c_i^eg_i^2 \right] ~,~\,
\end{eqnarray}
where  $y^u$, $y^d$ and $y^e$ are the Yukawa couplings
for the up-type quark, down-type quark, and lepton,
respectively. Also, $c^u$, $c^d$, and $c^e$ are given by 
\begin{eqnarray}
&& c^u=\left( \frac{13}{15}, 3, \frac{16}{3}\right) ~,~
c^d=\left( \frac{7}{15}, 3, \frac{16}{3}\right) ~,~
c^e=\left( \frac{9}{5}, 3, 0\right) ~.~\, 
\end{eqnarray}

\subsection{Beta Functions for the Vector-Like Particles}

We present one-loop and two-loop beta functions 
to the SM gauge couplings from the vector-like particles.
The general formulae are also given 
in Refs.~\cite{mac,Cvetic:1998uw,Barger:1992ac,Barger:1993gh,Martin:1993zk}.

First, we present the one-loop beta functions 
$\Delta b \equiv (\Delta b_1, \Delta b_2, \Delta b_3)$ as
complete supermultiplets from the extra particles
\begin{eqnarray}
\Delta b^{XQ + XQ^c} =({1\over 5}, 3, 2)~,~ \Delta b^{XU + XU^c} = ({8\over 5}, 0, 1)
~,~\Delta b^{XD + XD^c} = ({2\over 5}, 0, 1)~,~\,
\end{eqnarray}
\begin{eqnarray}
\Delta b^{XL + XL^c} = ({3\over 5}, 1, 0)~,~
\Delta b^{XE + XE^c} = ({6\over 5}, 0, 0)~,~
\Delta b^{XN + XN^c} = (0, 0, 0)~,~\,
\end{eqnarray}
\begin{eqnarray}
\Delta b^{XY + XY^c} =(5, 3, 2)~,~
\Delta b^{XT_i + \overline{XT}_i} = ({6\over 5}, 0, 0)~,~\,
\end{eqnarray}

Second, we present the  two-loop beta functions ($\Delta B_{ij}$) from
the extra particles in the non-supersymmetric models

\begin{eqnarray}
\Delta B^{XQ + XQ^c}=\pmatrix{\frac{1}{150}&
\frac{3}{10}&\frac{8}{15}\cr \frac{1}{10} & \frac{49}{2}& 8 \cr
\frac{1}{15}& 3 & \frac{76}{3} } ~,~
\Delta B^{XU + XU^c}=\pmatrix{\frac{64}{75}&
0 &\frac{64}{15}\cr 0 & 0 & 0 \cr
\frac{8}{15} & 0 & \frac{38}{3}} ~,~\,
\end{eqnarray}
\begin{eqnarray}
\Delta B^{XD + XD^c}=\pmatrix{\frac{4}{75}&
0 &\frac{16}{15}\cr 0 & 0 & 0 \cr
\frac{2}{15} & 0 & \frac{38}{3}} ~,~
\Delta B^{XL + XL^c}=\pmatrix{\frac{9}{50}&
\frac{9}{10}& 0 \cr \frac{3}{10} & \frac{49}{6}& 0 \cr
0 & 0 & 0 } ~,~\,
\end{eqnarray}
\begin{eqnarray}
\Delta B^{XE + XE^c}=\pmatrix{\frac{36}{25}&
0 & 0 \cr 0 & 0 & 0 \cr
0 & 0 & 0 } ~,~
\Delta B^{XN + XN^c}=\pmatrix{0&
0 & 0 \cr 0 & 0 & 0 \cr
0 & 0 & 0 } ~,~\,
\end{eqnarray}
\begin{eqnarray}
\Delta B^{XY + {{XY^c}}}=\pmatrix{\frac{25}{6}&
\frac{15}{2}&\frac{40}{3}\cr \frac{5}{2} & \frac{49}{2}& 8 \cr
\frac{5}{3}& 3 & \frac{76}{3} } ~,~
\Delta B^{XT_i + \overline{XT}_i}=\pmatrix{\frac{9}{25}&
0 & 0 \cr 0 & 0 & 0 \cr
0 & 0 & 0 }~.~\,
\end{eqnarray}

Third, we present the two-loop beta functions from
the extra particles in the supersymmetric models
\begin{eqnarray}
\Delta B^{XQ + XQ^c}=\pmatrix{\frac{1}{75}&
\frac{3}{5}&\frac{16}{15}\cr \frac{1}{5} & 21 & 16 \cr
\frac{2}{15}& 6 & \frac{68}{3} } ~,~
\Delta B^{XU + XU^c}=\pmatrix{\frac{128}{75}&
0 &\frac{128}{15}\cr 0 & 0 & 0 \cr
\frac{16}{15} & 0 & \frac{34}{3}} ~,~\,
\end{eqnarray}
\begin{eqnarray}
\Delta B^{XD + XD^c}=\pmatrix{\frac{8}{75}&
0 &\frac{32}{15}\cr 0 & 0 & 0 \cr
\frac{4}{15} & 0 & \frac{34}{3}} ~,~
\Delta B^{XL + XL^c}=\pmatrix{\frac{9}{25}&
\frac{9}{5}& 0 \cr \frac{3}{5} & 7 & 0 \cr
0 & 0 & 0 } ~,~\,
\end{eqnarray}
\begin{eqnarray}
\Delta B^{XE + XE^c}=\pmatrix{\frac{72}{25}&
0 & 0 \cr 0 & 0 & 0 \cr
0 & 0 & 0 } ~,~
\Delta B^{XN + XN^c}=\pmatrix{0&
0 & 0 \cr 0 & 0 & 0 \cr
0 & 0 & 0 } ~,~\,
\end{eqnarray}
\begin{eqnarray}
\Delta B^{XY + {{XY^c}}}=\pmatrix{\frac{25}{3}&
15 &\frac{80}{3}\cr 5 & 21 & 16 \cr
\frac{10}{3}& 6 & \frac{68}{3} } ~,~
\Delta B^{XT_i + \overline{XT}_i}=\pmatrix{\frac{18}{25}&
0 & 0 \cr 0 & 0 & 0 \cr
0 & 0 & 0 } ~.~\,
\end{eqnarray}

\section{ The Flipped $SU(5)\times U(1)_X$ models with  Vector-Like Particles}
\label{apdxB}

\subsection{ Renormalization Group Equations}

The two-loop renormalization group equations for the gauge couplings 
in the  flipped $SU(5)\times U(1)_X$ models~\cite{Ellis:1988tx} in the 
subsection A in Section II are
\begin{eqnarray}
(4\pi)^2\frac{d}{dt}~ g_i &=& b_i g_i^3 
 +\frac{1}{(4\pi)^2}
~ \sum_{j=1, 5} B_{ij} g_i^3 g_j^2 ~,~\,
\label{SUSYgauge}
\end{eqnarray}
where $i=1, 5$, and
the beta-function coefficients are 
\begin{eqnarray}
&&b=\left(\frac{15}{2},-5\right) ~,~~~  B=\pmatrix{\frac{33}{4}&
60\cr \frac{5}{2} & 82 }  ~.~ \\
\end{eqnarray}

\subsection{Beta Functions for the Vector-Like Particles}

First, we present the one-loop beta functions 
$\Delta b \equiv (\Delta b_1, \Delta b_5)$ as
complete supermultiplets from the extra particles
\begin{eqnarray}
\Delta b^{XF + \overline{XF}} =({1\over 2}, 3)~,~
\Delta b^{Xf + \overline{Xf}} = ({9\over 4}, 1)~,~\,
\end{eqnarray}
\begin{eqnarray}
\Delta b^{Xl + \overline{Xl}} =({5\over 4}, 0)~,~
\Delta b^{Xh + \overline{Xh}} = (1, 1)~,~\,
\end{eqnarray}
\begin{eqnarray}
\Delta b^{XT_i + \overline{XT}_i} =({5\over 4}, 0)~.~\,
\end{eqnarray}

Second, we present the two-loop beta functions from
the vector-like extra particles 
\begin{eqnarray}
\Delta B^{XF + \overline{XF}}=\pmatrix{\frac{1}{20}&
\frac{36}{5}\cr \frac{3}{10} & \frac{366}{5} }~,~
\Delta B^{Xf + \overline{Xf}} =\pmatrix{\frac{81}{40}&
\frac{108}{5}\cr \frac{9}{10} & \frac{98}{5} }~,~\,
\end{eqnarray}
\begin{eqnarray}
\Delta b^{Xl + \overline{Xl}} =\pmatrix{\frac{25}{8}&
0 \cr 0 & 0 }~,~
\Delta B^{Xh + \overline{Xh}} =\pmatrix{\frac{2}{5}&
\frac{48}{5}\cr \frac{2}{5} & \frac{98}{5} }~,~\,
\end{eqnarray}
\begin{eqnarray}
\Delta b^{XT_i + \overline{XT}_i} =\pmatrix{\frac{25}{32}&
0 \cr 0 & 0 } ~.~\,
\end{eqnarray}

\end{document}